\documentclass{aastex63}

\newcommand\lsim{\mathrel{\rlap{\lower4pt\hbox{\hskip1pt$\sim$}}
    \raise1pt\hbox{$<$}}}
\newcommand\gsim{\mathrel{\rlap{\lower4pt\hbox{\hskip1pt$\sim$}}
    \raise1pt\hbox{$>$}}}

\def\agnote#1{#1}

\submitjournal{ApJ}
\accepted{10/05/2020}

\shorttitle{A stream of hypervelocity stars from the Galactic Center}
\shortauthors{Generozov}
\graphicspath{{./}{figures/}}

\begin{document}
\correspondingauthor{Aleksey Generozov}
\email{alge9397@colorado.edu}

\title{A stream of hypervelocity stars from the Galactic Center}
\author{Aleksey Generozov}
\affiliation{JILA and the Astrophysical and Planetary Sciences Department\\
University of Colorado, Boulder}

\begin{abstract}
    Recent observations have found a 1700 km s$^{-1}$ star [S5-HVS1] that was ejected from the Galactic Center approximately five million years ago. This star was likely produced by tidal disruption of a binary. In particular, the Galactic Center contains a few million year old stellar disk that could excite binaries to nearly radial orbits via a secular gravitational instability. Such binaries would be disrupted by the central supermassive black hole, and would also explain the observed cluster of B stars $\sim 0.01$ pc from the Galactic Center. In this paper we predict S5-HVS1 is part of a larger stream, and use observationally motivated N-body simulations to predict its spatial and velocity distribution.
\end{abstract}
\keywords{Black Holes, Galaxy Nuclei, Hypervelocity Stars}

\section{Introduction}
\citet{hills1988} proposed that tidal disruptions of binary stars in the Galactic Center could produce hypervelocity stars--stars that are fast enough to escape the Galactic potential. In fact, \citet{koposov+2019} recently discovered S5-HVS1--a 1700 km s$^{-1}$ hypervelocity star. Tracing its path backwards in time, they found it would have passed through the Galactic Center 4.8 Myr ago. Although many other hypervelocity stars and candidates have been identified (see e.g. \citealt{brown+2005, brown+2014, boubert+2018} and the references therein), to date the \citet{koposov+2019} star has by far the strongest case for a Galactic Center origin.

Typically, in a binary disruption one of the stars is left bound to the SMBH, while the other is ejected from the Galactic Center. 
Binary disruptions in the Galactic Center would also explain the cluster of B-type stars with semimajor axes between $\sim0.005$ and $\sim0.05$ pc
(the ``S-stars''; \citealt{genzel+1997, ghez+1998, gillessen+2017}), as previously pointed out by \citet{ginsburg&loeb2006}, \citet{perets+2007}, \citet{lockmann+2009}, \citet{madigan+2009}, and \citet{dremova+2019}.

The flight time of S5-HVS1 is consistent with the age of the clockwise disk of stars between $\sim0.05$ to $\sim0.5$ pc from the Galactic Center (2.5-5.8 Myr; \citealt{lu+2013}). S5-HVS1's velocity vector is also consistent with a disk origin.
As discussed in \citet{madigan+2009} and \citet{generozov&madigan2020}, this disk could excite binaries to tidal disruption via a gravitational instability early in its evolutionary history. In particular, if this structure starts as an apsidally-aligned, eccentric disk some of its member stars and binaries can be excited to extreme eccentricities and tidally disrupted. However, the spectroscopic age of S5-HVS1 ($10^{7.72_{-0.33}^{+0.25}}$ years; \citealt{koposov+2019}) is greater than the disk's. This suggests it was a background star that was entrained in it.

In this scenario most disruptions would occur in a narrow range of times spanning a few$\times 10^5$ years (a few times the secular timescale of the disk). Also, the initial lopsided geometry of the disk would be imprinted on the spatial distribution of ejected stars. In particular, this lopsidedness is apparent in the orientation of orbits that pierce the binary tidal disruption radius in the disk simulations of \citet{generozov&madigan2020}. Such stars occupy a relatively narrow solid angle, so binary disruption would eject stars in a slender cone. In summary, we predict the Galactic Center produced a conical stream of high velocity stars $\sim5$ Myr ago.

The galactocentric distance of stars in this cone would be set by their ejection velocity. In this paper, we quantify the expected properties of this structure (the ``Galactic Center Cone'' or GCC stream) to facilitate a search in \emph{Gaia} data. 
Our predictions include not only hypervelocity stars, but also somewhat slower stars still bound to the Galaxy.  
Previous work has found that an eccentric intermediate mass black hole (IMBH) can also produce a broad cone of stars. The cone will be aligned with the IMBH's velocity vector at pericenter, which would shift over time as the IMBH orbit precesses \citep{levin2006, sesana+2008}. However, the existence of an IMBH in the Galactic Center is severely constrained by existing observations (see \citealt{gravity+2020} and the references therein). Finally, anisotropic bursts of hypervelocity stars can be produced by the close encounter of a star cluster with the Galactic Center (see Figure 5 in \citealt{fragione+2017}).

\section{The GCC stream} 
We use S5-HVS1 to anchor our prediction. In our scenario, the GCC stars are ejected in a narrow range of times from an apsidally-aligned disk, which implies that stars are ejected in a cone near this star's velocity vector. We use the N-body simulations of eccentric disks in the Galactic Center from \citet{generozov&madigan2020} to constrain the geometry of this cone.

In these simulations, we initialized disks of 300 equal mass particles in an eccentric, apsidally-aligned 
configuration. The total disk mass is $1-2 \times 10^4 M_{\odot}$. The particles have a range of semimajor axes that is similar to the present-day clockwise disk. The disk orbits a $4\times 10^6 M_{\odot}$ SMBH, and evolves due to its own self-gravity and the influence of a spherical potential (representing the old stellar population that contains most of the stars in the Galactic Center). These simulations also include an approximate treatment of general relativistic effects. Stellar binaries are not explicitly included in the simulations, but we record binary disruptions when particles pass within $3\times 10^{-4}$ pc of the central SMBH. (This is a representative tidal radius for binary stars). 

\begin{figure}
    \centering
    \plottwo{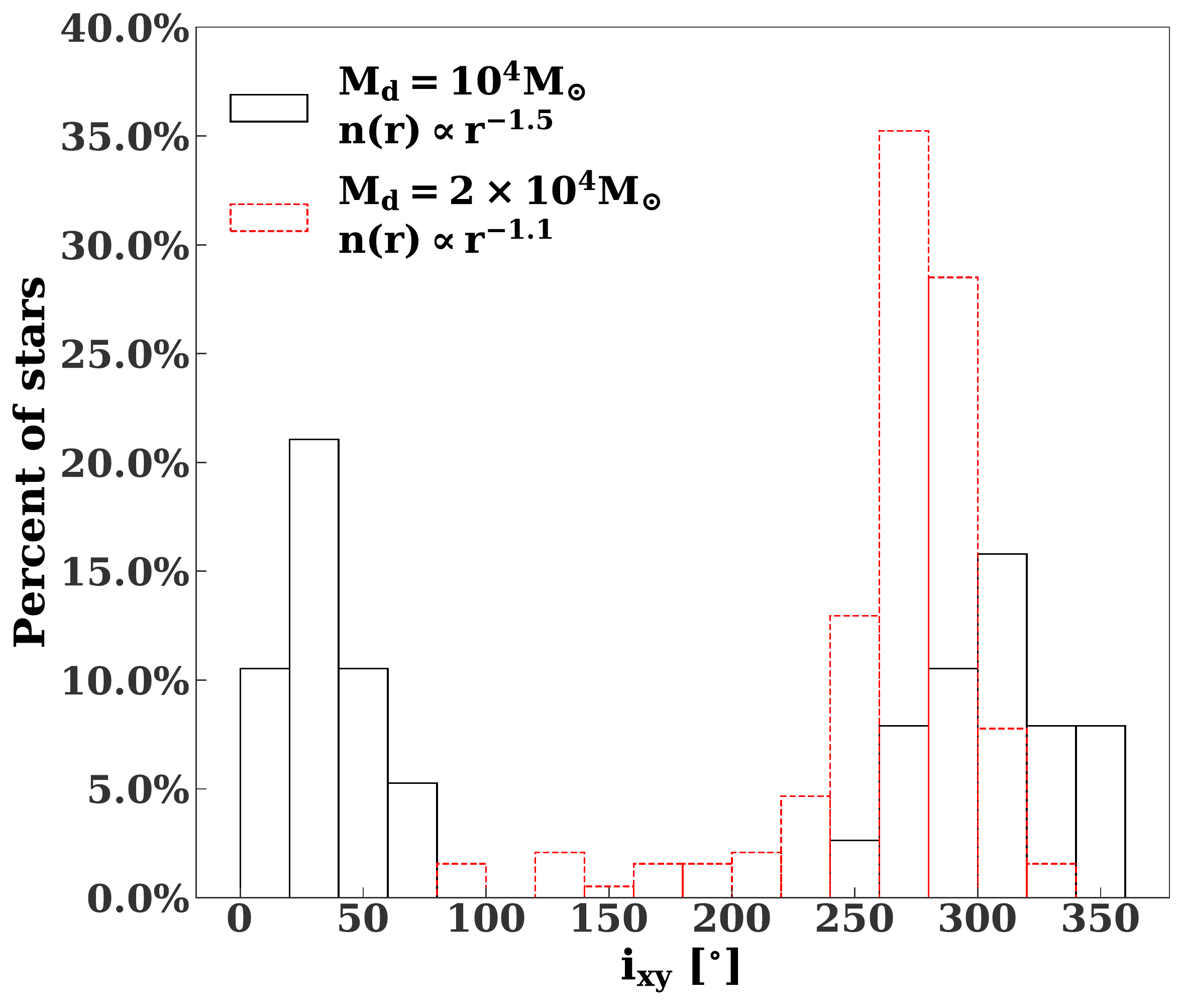}{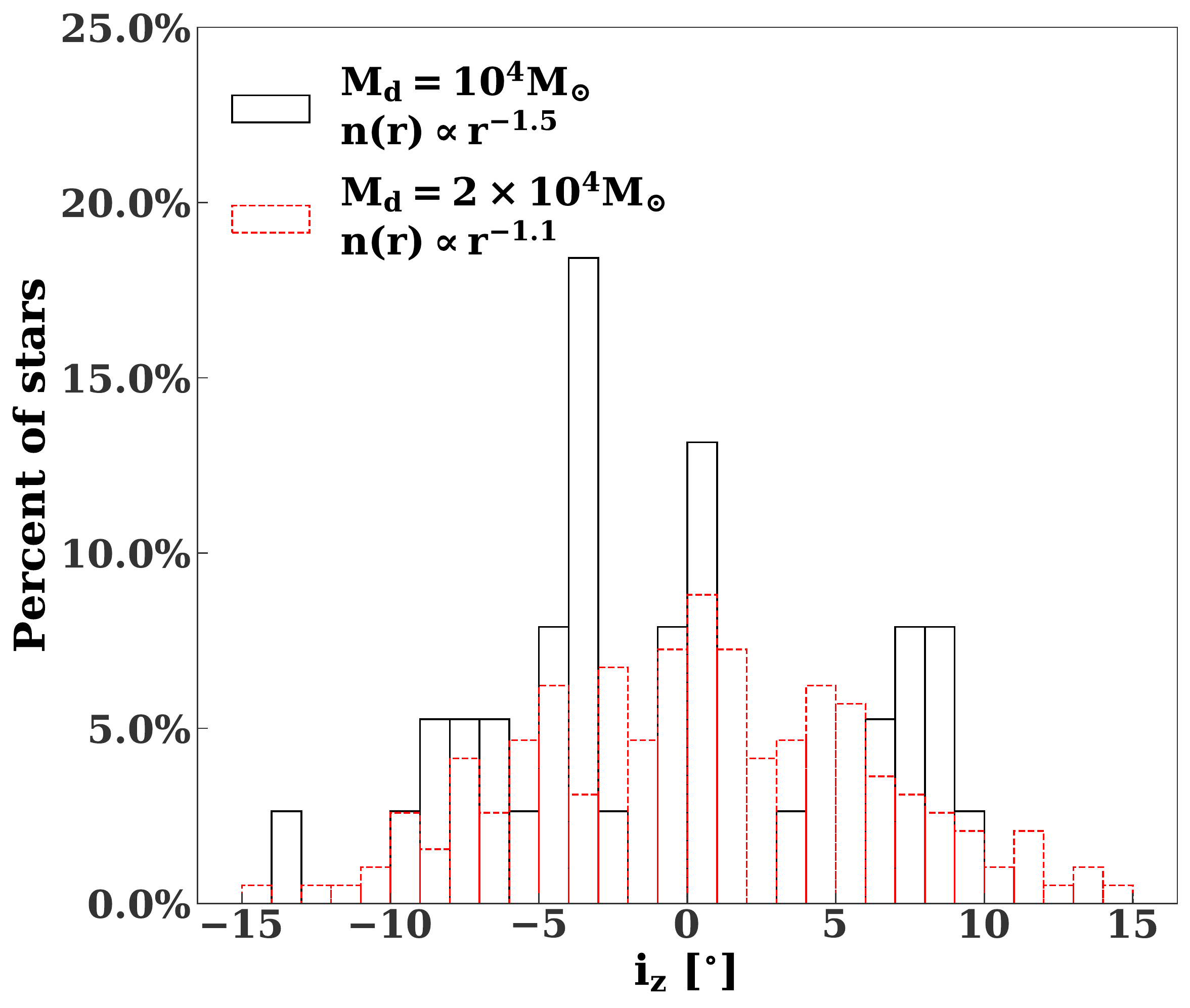}
    \caption{\emph{Left panel:} Histogram of $i_{xy}$ for ejected stars in our disk simulations (different line styles correspond to different disk masses and background stellar potentials). This is the angle between the disk plane projection of an ejected star's velocity and a reference axis. Each histogram is constructed by stacking the results of several simulations. \emph{Right panel:} Histogram of $i_z$ for these simulations. This is the angle between an ejected star's velocity and the disk plane at infinity. For these distributions, we exclude any disruptions after the first Myr.}
    \label{fig:ie}
\end{figure}

The orbital orientations of disrupted binaries will determine the angular spread of the ejected stars' velocity vectors. (The binary's orbital motion will be a second-order effect, as the center of mass velocity is $\sim$100 times larger than the binary's internal velocity at disruption). To quantify this spread, we assume each ejected star is on a hyperbolic orbit with the same pericenter, inclination, longitude of ascending node, and argument of pericenter as the progenitor binary.
The left panel of Figure~\ref{fig:ie} shows the distribution of $i_{\rm xy}$, the angle between the velocity vectors projected into the disk plane and a reference axis, in our simulations. The circular standard deviation of $i_{\rm xy}$ for the red (black) distribution is $35^{\circ}$ ($53^{\circ}$). The right panel of Figure~\ref{fig:ie} shows the distribution of $i_z$, the angle between the velocity vectors and the disk plane. The circular standard deviation of $i_z$ for the red and black distributions is $6^{\circ}$. Although the number of particles in our simulations is likely a factor of a few smaller than the real disk \citep{lu+2013}, we find that the spread of these angles is a weak function of the number of stars in the disk. Decreasing the particle number from 300 to 100, increases the circular standard deviation of $i_{xy} $ by 3$^{\circ}$ (for the red simulations) and 14$^{\circ}$ (for the black simulations).

We simulate close binary-SMBH encounters with \texttt{AR--Chain} to obtain the velocity distribution of stream stars. In these encounters the binary stars are (independently) drawn from an $m^{-1.7}$ mass function extending from $1$ to $60 M_{\odot}$. The semimajor axis is drawn from a log-uniform distribution. The minimum and maximum binary semimajor axis are functions of the binary mass and eccentricity, and are set by the prescriptions in $\S$~4 of \citet{generozov&madigan2020} (the semimajor axis is typically between a few$\times 10^{-2}$ and a few au). The binaries' centers of mass are on eccentric orbits with a semimajor axis of 0.05 pc (approximately the inner edge of the clockwise disk). The pericenter of each binary's center of mass orbit is its effective tidal radius,

\begin{equation}
    r_{\rm t, eff}=\chi(e_{\rm bin}) \left(\frac{M}{m_{\rm bin}}\right)^{1/3} a_{\rm bin},
\end{equation}
where $M$ is the mass of the SMBH; $m_{\rm bin}$, $a_{\rm bin}$, and $e_{\rm bin}$ are the mass, semimajor axis, and eccentricity of the binary respectively; $\chi$ is a factor of order unity (see $\S$~4 of \citealt{generozov&madigan2020} for details). This distribution of binary properties approximately reproduces the observed semimajor axis distribution of the S-stars (see Figures 6, 7, and surrounding discussion in \citealt{generozov&madigan2020}).\footnote{The distribution of binary properties in these simulated encounters is similar to the input distribution for Figure 7 of \citet{generozov&madigan2020}. However, the mass function is broader here.}

The number of stars in the stream can be constrained from the observed number of S-stars. There are at least 22 S-stars with semimajor axes less than 0.03 pc.\footnote{We focus on stars with smaller semimajor axes to avoid contamination from stars that may have been kicked out of the disk by other mechanisms (e.g. by vector resonant relaxation; \citealt{szoelgyen&kocsis2018}).} The stars likely have masses between $\sim 3$ and $15 M_{\odot}$, considering the spectroscopic mass measurements in \citet{habibi+2017} and their K-band magnitudes \citep{cai+2018}. Higher mass stars are notably absent from the S-stars, although they are present in the disk. This can be understood at least qualitatively in the Hills model. For nearly parabolic disruptions the primary and secondary are equally likely to be left bound to the SMBH, but the primary would be deposited at larger semimajor axis \citep{kobayashi+2012, generozov&madigan2020}. Assuming that the masses of the observed S-stars are drawn from the $m^{-1.7}$ mass function of the disk \citep{lu+2013} with a truncation at low mass due to observability and a truncation at higher masses due to the above mass ratio effect, there would be at least $\sim$60 stars in the S-star cluster above $1 M_{\odot}$. In reality, the mass function of the S-stars would not have a sharp truncation, but would gradually steepen towards higher masses. The post-disruption mass function of bound stars can be fit with a ``Nuker profile'' (typically used for fitting galaxy surface brightness profiles; see e.g. \citealt{lauer+2005}), viz.

\begin{equation}
    f(m)=k \left(\frac{m}{m_b}\right)^{-\gamma} \left( 1 +  \left(\frac{m}{m_b}\right)^{\alpha}  \right)^{(\gamma-\beta)/\alpha}.
\end{equation}
This profile approaches $m^{-\gamma}$  ($m^{-\beta}$) for masses much less (greater) than $m_b$. The smoothness of the transition is controlled by $\alpha$, and $k$ is an overall normalization factor. The maximum likelihood parameters for stars deposited inside of 0.03 pc in our simulated binary disruptions are $\{m_b, \gamma, \beta\}=\{2.5  M_{\odot}, 2, 2.4\}$ for fixed $\alpha=10$. This mass function suggests there are at least $\sim$90 S-stars with masses greater than $1 M_{\odot}$.
 
We construct a mock stream by selecting 90 simulated encounters where (a) the binary is tidally separated into two stars and (b) one of the stars has a semimajor axis less than 0.03 pc. The mock stream's axis is perpendicular to the angular momentum of the clockwise disk, and aligned with the disk plane projection of S5-HVS1's velocity vector.\footnote{Here the unit angular momentum vector of the disk is $\langle 0.785,0.196,-0.588 \rangle$ (based on Figure 12 in \citealt{gillessen+2017}), and the velocity vector of S5-HVS1 is $\langle -786,-381,-1570 \rangle$ km s$^{-1}$ (based on Figure 6 in \citealt{koposov+2019}) in the Galactic Standard of Rest.} To obtain the direction of travel of each stream star, we rotate the stream axis about the angular momentum of the clockwise disk by a random angle $\theta_{xy}$, and then rotate the resulting vector out of the disk plane by another random angle $\theta_z$. We assume that $\theta_{xy}$ and $\theta_z$ are normally distributed, with standard deviations of 35 and 6$^{\circ}$ respectively (the circular standard deviations of the red distributions in Figure~\ref{fig:ie}). 
 
The angle between an ejected star's velocity at disruption and its velocity at infinity is a function of pericenter. In our disk simulations the pericenter distribution is artificially narrow, as all particle have the same tidal radius. In reality, binaries would have a broad range of semimajor axes, and thus a broad range of tidal radii. In general, the deflection angle between pericenter and infinity is (see e.g. Chapter 5 of \citealt{merritt2013})

\begin{eqnarray}
    \xi &=& \frac{1}{2} \arctan{\frac{2 p/p_o}{(p/p_o)^2-1}}  \nonumber \\
    p_o  &=& \frac{G M}{v_{\rm ej}^2},
\end{eqnarray}
where $M$ is mass of the SMBH, $v_{\rm ej}$ is the velocity of the star after escaping the SMBH potential, and $p$ is the impact parameter of the encounter. From $\S$~5 of \citet{generozov&madigan2020}
\begin{eqnarray}
    \frac{p}{p_o} &\approx& x \left(1+ \frac{2}{x} \right)^{1/2} \nonumber \\ 
    x             &\equiv& \left(\frac{M}{m_{\rm bin}}\right)^{-1/3},
\end{eqnarray}
for parabolic encounters.  For $m_{\rm bin}=20 M_{\odot}$, $\xi\approx 79^{\circ}$. In our simulated binary-SMBH encounters, $\xi$ ranges from $67^{\circ}$ to $89^{\circ}$.  Thus, the scatter in the deflection angle is smaller than the spread in $i_{xy}$ (see Figure~\ref{fig:ie}) and can be neglected.

\begin{figure}
    \centering
    \plottwo{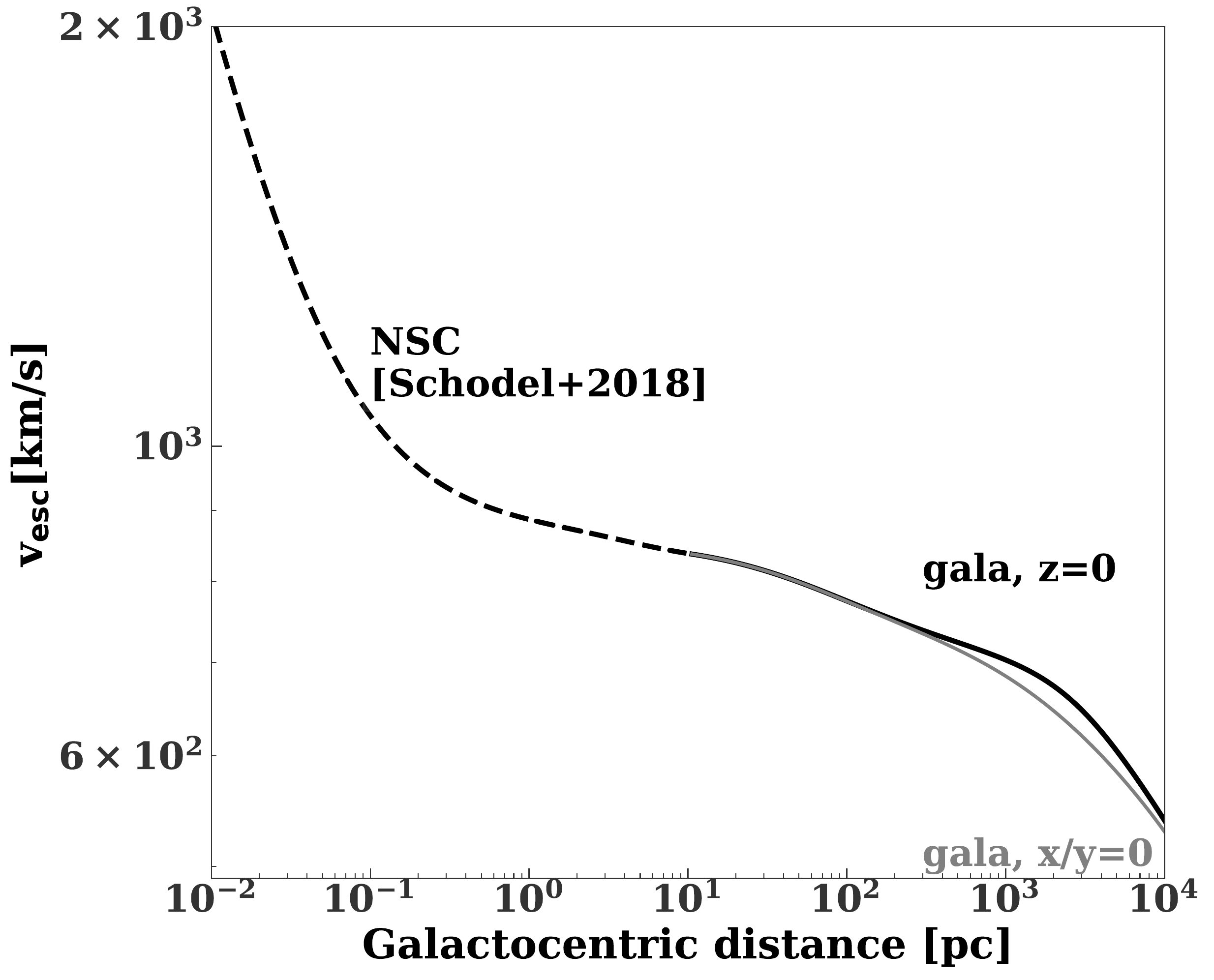}{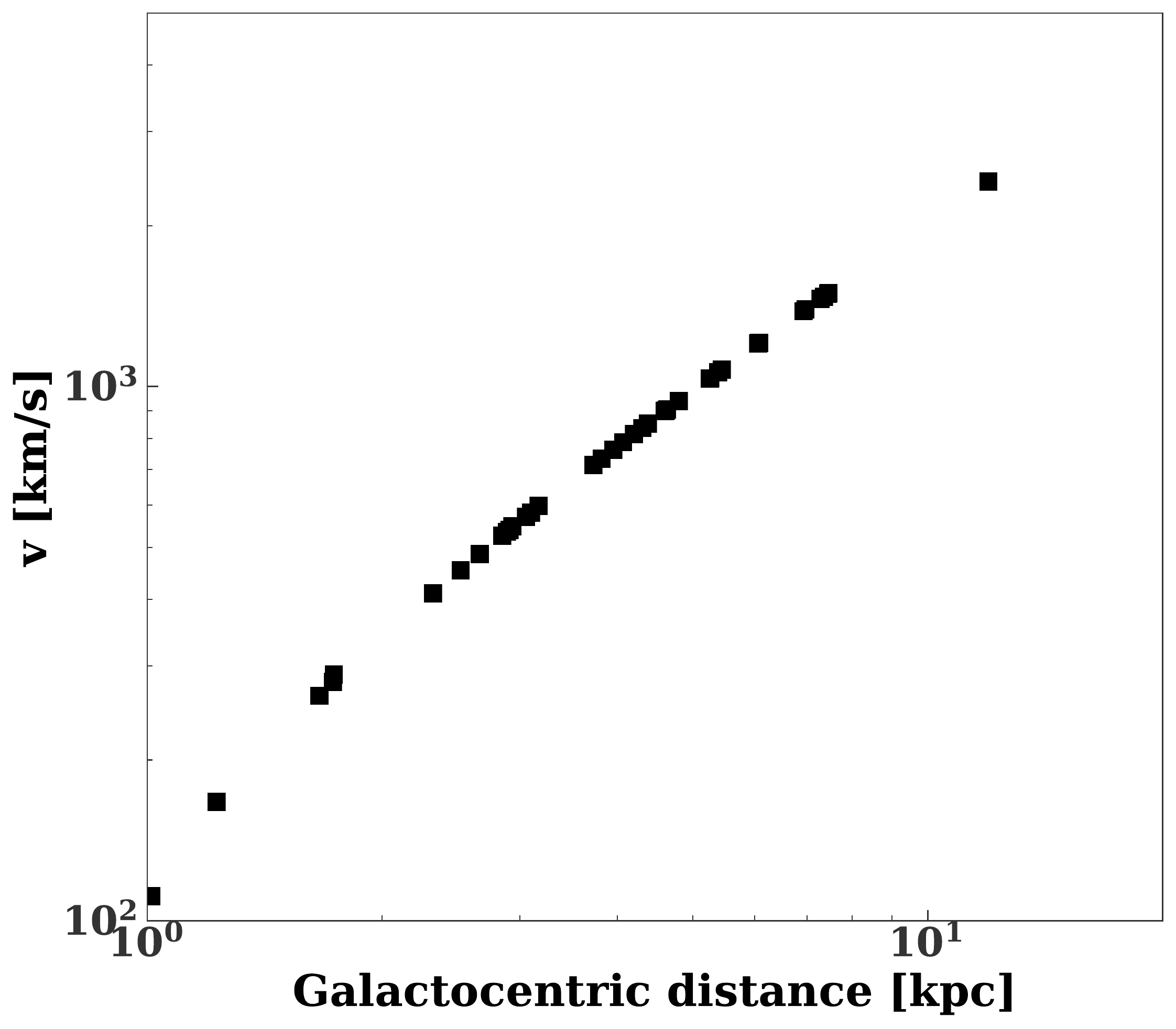}
    \caption{\emph{Left panel:} Escape velocity as a function of Galactocentric distance for our assumed potential. The solid lines are from the default Milky Way potential model in \texttt{gala}, which is based on recent mass measurements between 10 pc and 150 kpc. The gray (black) line shows the escape velocity moving perpendicular (parallel) to the Galactic disk. The dashed line shows the contribution of the Galactic Center SMBH and nuclear star cluster (based on fits to the Galactic Center density profile in \citealt{schodel+2018}). \emph{Right panel:} Velocity versus galactocentric distance in our mock stream after 4.8 Myr. The stream stars are all ejected from the Galactic Center at the same time, so there is a one-to-one relation between distance and velocity. The slowest stars in the stream are significantly decelerated by the Galactic potential.}
    \label{fig:rv}
\end{figure}

Finally, we integrate each star's orbit through a Milky Way potential model for 4.8 Myr using the \texttt{gala} software package \citep{gala}. In particular, we use the default ``MilkyWayPotential'' class, which contains a Hernquist bulge and nucleus, a Miyamoto-Nagai disk, and an NFW dark matter halo. The parameters for each of these components are derived by fitting to a set of recent mass measurements between 10 pc and 150 kpc (the disk and bulge parameters are the same as those in \citealt{bovy2015}; the full list of mass measurements used to constrain the halo properties are drawn from various sources).\footnote{The full list can be found at \url{https://gala-astro.readthedocs.io/en/latest/potential/define-milky-way-model.html\#Introduction}.} The enclosed mass at 10 pc in this model is anchored by \citet{feldmeier+2014} to $3\times 10^7 M_{\odot}$, while the enclosed mass at 120 pc is anchored by \citet{launhardt+2002} to $8\times 10^8 M_{\odot}$ (the uncertainties in these mass measurements are $\sim 10\%$). The solid lines in the left panel of Figure~\ref{fig:rv} show the escape velocity as a function of galactocentric radius for the \texttt{gala} model, while the dashed line shows the contribution of a $4\times 10^6 M_{\odot}$ SMBH and the best fit for the stellar density profile inside of 10 pc from \citet{schodel+2018} (normalized so that the total mass inside of 10 pc is $3\times 10^7 M_{\odot}$). 

\begin{figure}
    \centering
    \gridline{\fig{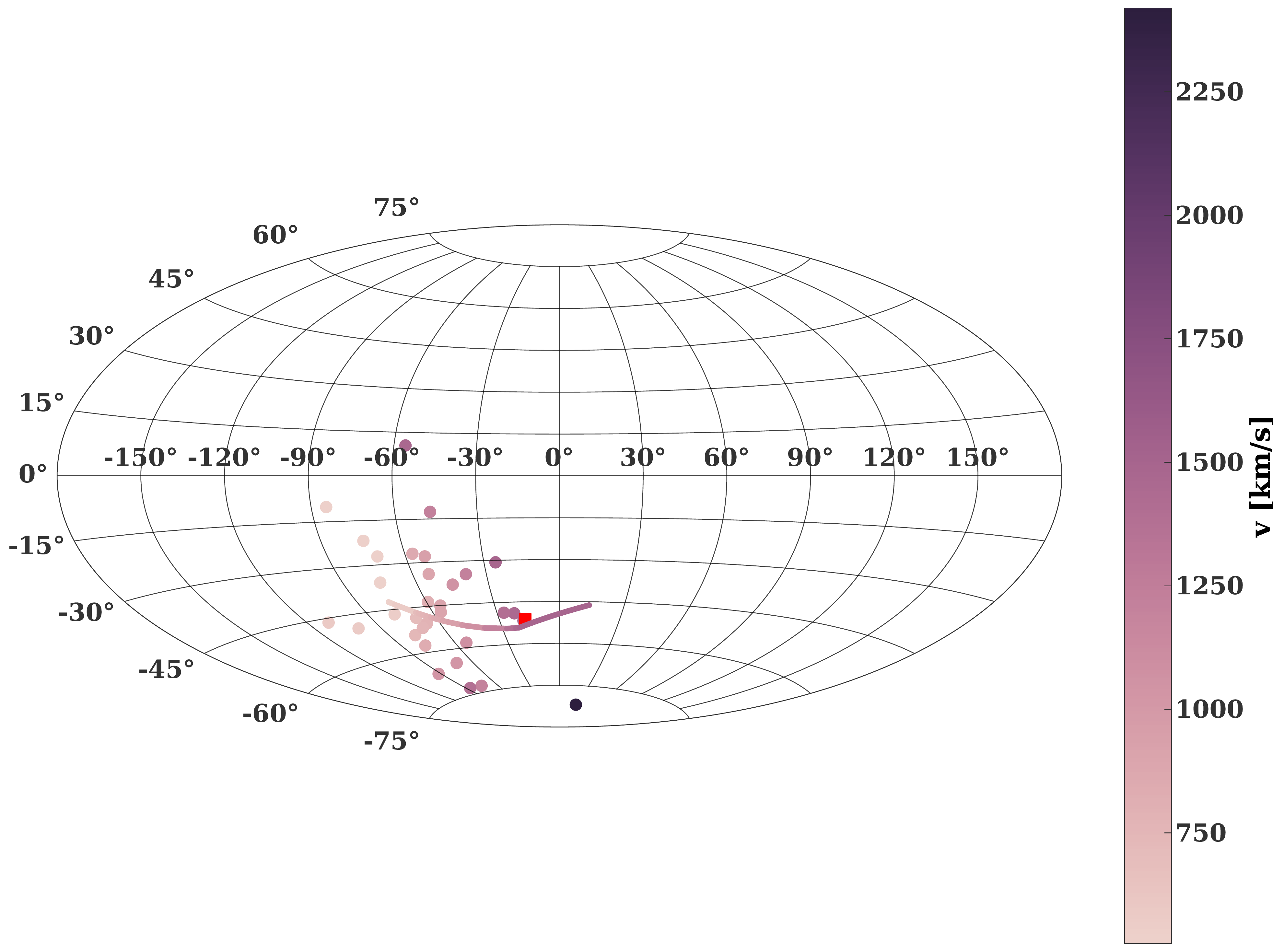}{0.7\textwidth}{}}
    \gridline{\fig{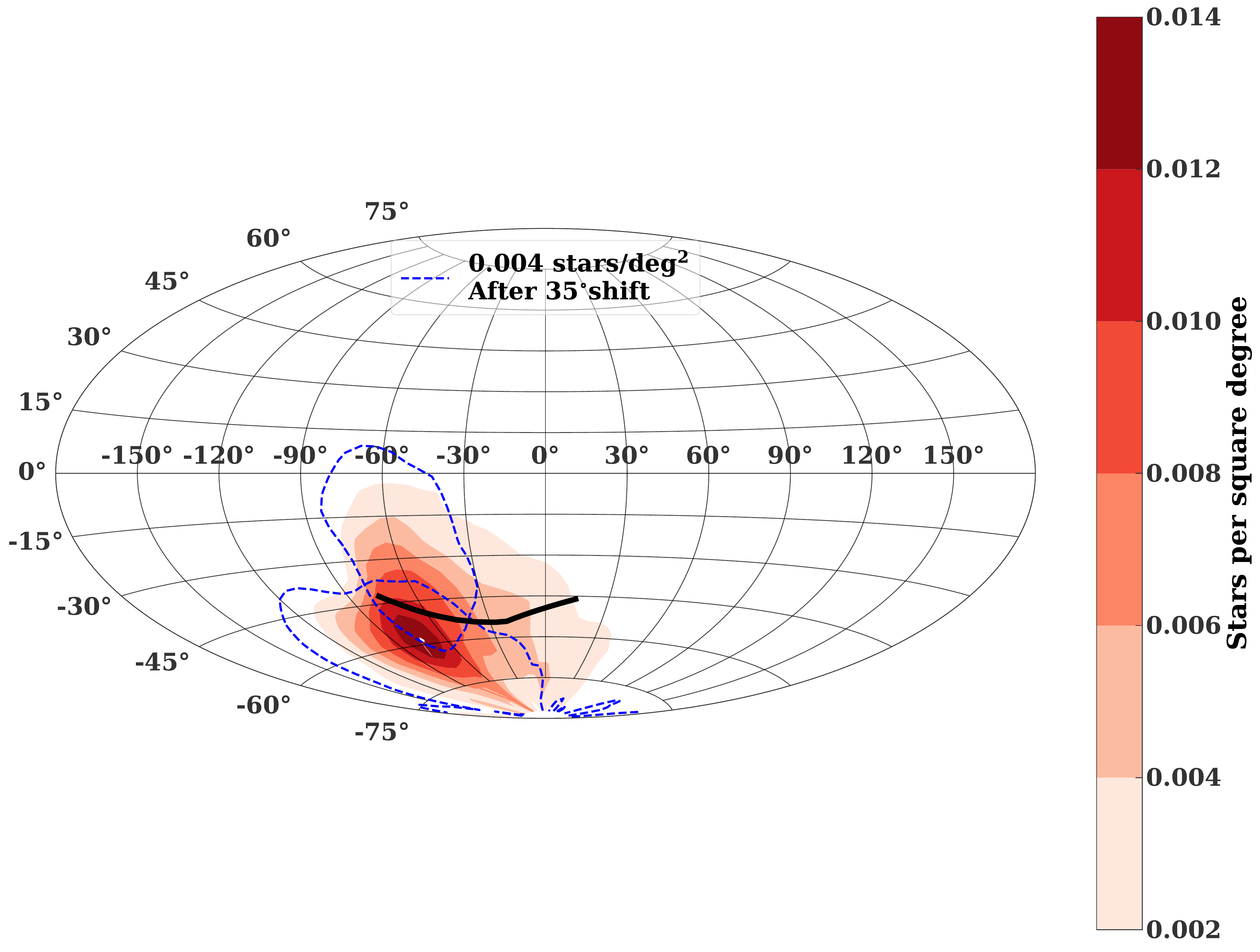}{0.7\textwidth}{}}
    \caption{\emph{Top panel:} Aitoff projection of stars in our mock stream with velocities of at least 500 km s$^{-1}$ after deceleration by the Galactic potential (in total 32 stars). The coordinates are right ascension and declination. The line is where the stars would end up if they were all ejected along the axis of the stream. The red square is S5-HVS1. \emph{Bottom panel:} The red contours show the density of stream stars on the sky. This plot is constructed by stacking many realizations of our mock stream. \agnote{The dashed, blue lines show how the contour corresponding to 0.004 stars per square degree shifts, as the stream axis is rotated by $35^{\circ}$ in either direction (placing the S5-HVS1 near the edge of the stream).}}
    \label{fig:aitoff}
\end{figure}

Each stream star is initialized at 10 pc with velocity
\begin{equation}
    v_o = \sqrt{2 \left(e_o-\Phi_o\right)},
\end{equation}
where $e_o$ is the specific energy of the star, and $\Phi_o$ is the potential energy at 10 pc (including the nuclear star cluster). The bottom panel of Figure~\ref{fig:rv} shows the final velocity of ejected stars as a function of galactocentric distance. The stream stars are mostly decelerated by the galactic bulge, which acts as high pass filter: only stars with velocities $\gsim$ 500 km s$^{-1}$ (after escaping the SMBH and nuclear star cluster potential) reach 1 kpc (see also the review by \citealt{brown2015}). Figure~\ref{fig:aitoff} shows an Aitoff projection of the fastest stream stars (with velocities of at least 500 km s$^{-1}$ after 4.8 Myr). 
The red square is S5-HVS1 \citep{koposov+2019}. The stream stars are concentrated in a relatively small region of the sky, and there is a strong correlation between a star's position along the stream and its velocity.  

So far we have assumed that the axis of the stream is aligned with S5-HVS1. However, this is not necessarily the case. The blue dashed lines in the bottom panel of Figure~\ref{fig:aitoff} shows how the distribution of stars would shift if the stream axis is rotated by $35^{\circ}$ in either direction (so that S5-HVS1 is closer the edge of the stream).

We find no strong correlation between stellar mass and velocity within the stream. The overall stream mass function is close to $m^{-2}$.

\section{Summary}
Recent observations suggest that some binaries from the clockwise disk in the Galactic Center were tidally disrupted five million years ago. These binary disruptions would typically leave one star bound to the central SMBH (like the S-stars), and one star ejected towards infinity (like S5-HVS1; \citealt{koposov+2019}). In this paper, we predict that Galactic Center produced a cone of high velocity stars (the ``GCC'') in the recent past. To facilitate a search in \emph{Gaia} data, we quantify the spatial and velocity distributions of stars in the GCC using S5-HVS1 and observationally motivated N-body simulations. In particular, the stars have galactocentric distances of a few to 10 kpc, and are in a cone structure with an opening angle of $\sim 30^{\circ}$.

\acknowledgements
{
I thank the anonymous referee for a constructive report. I thank Elena Rossi for suggesting this project at the YITP-T-19-07 International
Molecule-type Workshop, ``Tidal Disruption Events: General Relativistic Transients.''
I thank Ann-Marie Madigan and Jason Dexter for helpful comments and insightful conversations.

I acknowledge support from NASA Astrophysics Theory Program (ATP) under grant NNX17AK44G 

This work utilized resources from the University of Colorado Boulder Research Computing Group, which is supported by the National Science Foundation (awards ACI-1532235 and ACI-1532236), the University of Colorado Boulder, and Colorado State University. 
}

\software{\texttt{AR--Chain} \citep{mikkola&merritt2008}, \texttt{REBOUND} \citep{rein.liu2012}, 
\texttt{REBOUNDX} \citep{tamayo+2019}, \texttt{gala} \citep{gala}
\texttt{AstroPy} \citep{astropy+2018}, \texttt{Matplotlib} \citep{hunter+2007}, \texttt{NumPy}, \texttt{SciPy} \citep{2020SciPy-NMeth}, IPython \citep{perez+2007}}

\clearpage
\footnotesize{
\bibliographystyle{aastex}
\bibliography{master}
}

\end{document}